# Improvement of Applicability in Student Performance Prediction Based on Transfer Learning

Yan Zhao*
Beijing Polytechnic
Beijing, China
*zhaoyan3@bpi.edu.cn

Pengbo Wang
Beijing Information Science and Technology University
Beijing, China

Hao Wang
The Sixth Research Institute
China Electronics Corporation
Beijing, China
wanghao8888@outlook.com

*Abstract*—Predicting student performance under varying data distributions is a challenging task. This study proposes a method to improve prediction accuracy by employing transfer learning techniques on the dataset with varying distributions. Using datasets from mathematics and Portuguese language courses, the model was trained and evaluated to enhance its generalization ability and prediction accuracy. The datasets used in this study were sourced from Kaggle, comprising a variety of attributes such as demographic details, social factors, and academic performance. The methodology involves using an Artificial Neural Network (ANN) combined with transfer learning, where some layer weights were progressively frozen, and the remaining layers were fine-tuned. Experimental results demonstrated that this approach excels in reducing Root Mean Square Error (RMSE) and Mean Absolute Error (MAE), while improving the coefficient of determination ($R^2$). The model was initially trained on a subset with a larger sample size and subsequently fine-tuned on another subset. This method effectively facilitated knowledge transfer, enhancing model performance on tasks with limited data. The results demonstrate that freezing more layers improves performance for complex and noisy data, whereas freezing fewer layers is more effective for simpler and larger datasets. This study highlights the potential of transfer learning in predicting student performance and suggests future research to explore domain adaptation techniques for unlabeled datasets.

I. INTRODUCTION

The current educational domain is undergoing significant innovations and transformations driven by big data technologies, making student performance prediction an increasingly important research topic. Accurately predicting student performance not only assesses students' academic mastery but also provides valuable reference indicators for schools, parents, and teachers. This information aids in adjusting teaching plans, providing personalized tutoring, and ultimately improving the overall quality of education and student outcomes. However, traditional methods for predicting student performance, such as linear regression and other statistical techniques [1, 2], while capable of reflecting student learning trends to some extent, suffer from significant limitations in terms of prediction accuracy and model generalizability. These methods predominantly rely on simple linear relationships and are inadequate in capturing the complexity of student learning behaviors and the diversity of data features. Consequently, their performance is often suboptimal, particularly when dealing with multifaceted and heterogeneous educational data.

In response to these limitations, the advancement of Artificial Intelligence (AI) technologies, especially the progress in machine learning methods, offers new opportunities for improving student performance prediction due to their superior prediction ability in many fields such as automatic driving, healthcare and industries [3-5]. Researchers have begun to explore how these advanced technologies can be utilized to achieve more accurate predictions, thereby better meeting the needs of personalized education [6]. For instance, some studies have employed neural networks, support vector machines, and other machine learning algorithms for predicting student performance, and demonstrated relatively favorable results [7, 8]. Neural networks, with their ability to model complex, non-linear relationships, and support vector machines, with their robustness in high-dimensional spaces, have both shown promise in enhancing prediction accuracy. Additionally, other research efforts have explored the use of ensemble learning methods e.g. decision trees and random forests [9, 10]. These methods have demonstrated advantages in handling complex datasets by combining the predictions of multiple models to improve overall performance. Despite these advancements, a common challenge that persists across these methodologies is their limited generalizability and applicability to diverse data distributions. Models trained on one dataset often perform poorly when applied to different datasets with varying distributions, highlighting the need for more adaptable and robust approaches. Furthermore, the dynamic nature of student populations presents additional challenges. Each academic year brings new cohorts of students with varying characteristics and learning behaviors. Consequently, models developed for one cohort may quickly become outdated, necessitating continuous data collection and model retraining. This increases both operational complexity and resource requirements for maintaining predictive accuracy. Additionally, student performance data often vary significantly due to changes in curriculum, teaching methods, and other contextual factors, further complicating the prediction task. To address these challenges, transfer learning can be considered as a promising technique. It involves transferring knowledge from one domain or task to another, thereby enhancing model performance on tasks with limited data. This approach can effectively mitigate the issue of performance degradation caused by differences in data distributions [11]. Transfer learning has achieved significant success in various fields [12, 13], such as image processing and natural language processing, by leveraging pre-trained models to improve accuracy and efficiency. However, its application to student performance prediction remains underexplored in the current literature.

In light of this research gap, this study proposes a novel student performance prediction method that integrates transfer learning with K-means clustering. The primary objective is to enhance the model's applicability and accuracy across different data distributions. The proposed method involves several key steps. First, the K-means clustering algorithm is used to partition the dataset into multiple subsets, each characterized by distinct data distribution patterns. This step is crucial for identifying clusters of students with similar learning behaviors and performance characteristics. Subsequently, an initial predictive model is trained on the subset with the largest data volume. This model serves as the foundation for subsequent predictions. Considering that models trained on specific subsets may not be able to generalize well to other subsets with different distributions, in order to further improve predictive performance, transfer learning techniques with weight adjustment and parameter optimization were introduced. The initial model was adjusted using transfer learning techniques to adapt to other subsets, improving the applicability and accuracy of the model on different data distributions.

## II. METHOD

### A. Dataset collection

The dataset for this study was sourced from Kaggle [14], a well-established platform known for hosting data science competitions and providing a collaborative environment for data scientists and machine learning practitioners.

This study utilizes two distinct but related datasets from Kaggle, focusing on student performance in Mathematics and Portuguese language courses at two Portuguese secondary schools. Data were collected through school reports and questionnaires, capturing a wide range of attributes such as demographic details (e.g., age, sex), social factors (e.g., parental education, family size), and academic performance metrics (e.g., study time, absences number of school absences ).It contains first period grade G1,second period grade G2,final grade G3.The Mathematics dataset comprises 395 entries, and the Portuguese language dataset includes 649 entries, with each entry characterized by 33 attributes. Although the dataset includes grades for three periods (G1, G2, and G3), G3 is selected as the target variable for prediction because it represents the cumulative performance of students and provides a comprehensive assessment of their academic achievement.

### B. Dataset preprocessing

The preprocessing of the dataset involved converting categorical variables to numerical formats to facilitate machine learning analysis and ensure compatibility with various predictive models. In the original dataset, some attributes were recorded in textual format. To streamline the model processing, these textual attributes were converted into numerical formats. For instance, in the Fjob (father's job) field, "teacher" and "other" were encoded as 4 and 2, respectively. Similarly, other textual attributes were also converted to numerical formats to ensure data consistency and processability.

To produce subsets with different distributions for evaluating the applicability of models, this study employed the K-means clustering algorithm [15]. K-means is an unsupervised learning algorithm used to partition a dataset into k clusters, maximizing the similarity within each cluster while maximizing the difference between clusters. The feature matrix X was utilized for K-means clustering to ensure diversity in data distribution. To determine the optimal number of clusters (K), the study resorted to the Elbow algorithm [16]. This method involves calculating the Within-Cluster Sum of Squares (WCSS) for different values of k and plotting the relationship between k and WCSS. By observing the plot as shown in Figure 1, it can be noted that as k increases, the WCSS decreases at a diminishing rate, forming a distinct bend (elbow) at a certain point. Based on this method, the optimal K value was determined to be 2 for both datasets, as the gradient is highest at this point, indicating that dividing the dataset into two clusters is the most reasonable.

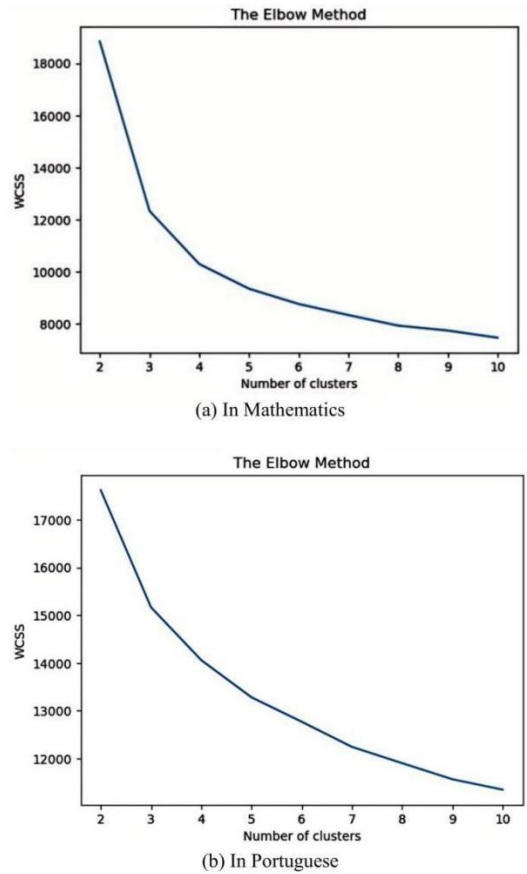

Figure 1. The optimal number of clusters based on Elbow Method.

For visualization purposes and to intuitively observe data distribution and clustering effects, this study employed Principal Component Analysis (PCA) [17] to reduce the dimensionality of the data to two dimensions. PCA is a statistical technique that transforms the original variables into a new set of orthogonal variables called principal components, which capture the maximum variance in the data. This method helps in simplifying the complexity of high-dimensional data while retaining its essential patterns. The visualized results of the reduced data are shown in Figure 2.

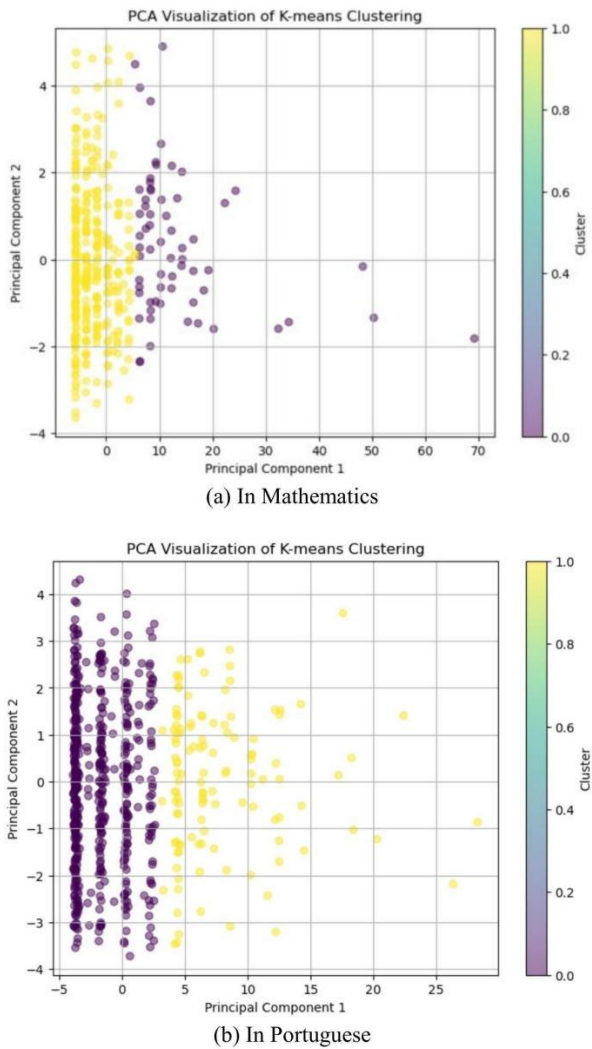

Figure 2. PCA Visualization of K-means Clustering.

In the data preprocessing phase, the dataset was divided into a training set and a test set, with 70% of the data used for training and 30% for testing. This splitting method facilitates the evaluation of the model's performance on unseen data, thereby providing a better assessment of its generalization capability.

Additionally, to accelerate the convergence of model training, normalization was applied to the data. Normalization involves scaling features to a common range, typically between 0 and 1. By standardizing feature values, the differences in magnitudes between features are minimized, enhancing the efficiency and stability of optimization algorithms. This process prevents extreme feature values from adversely affecting the training process, thereby improving the efficiency and performance of the model training.

### C. Transfer learning-based ANN model

#### 1) Introduction of ANN model

The fundamental concept of Artificial Neural Networks (ANN) is to simulate the structure of human brain neurons, utilizing input layers, hidden layers, and output layers to perform complex high-level representation learning illustrated in Figure 3. Each layer of the neural network consists of numerous neurons, which transmit information through weighted connections, emulating the learning and decision-making processes of the brain.

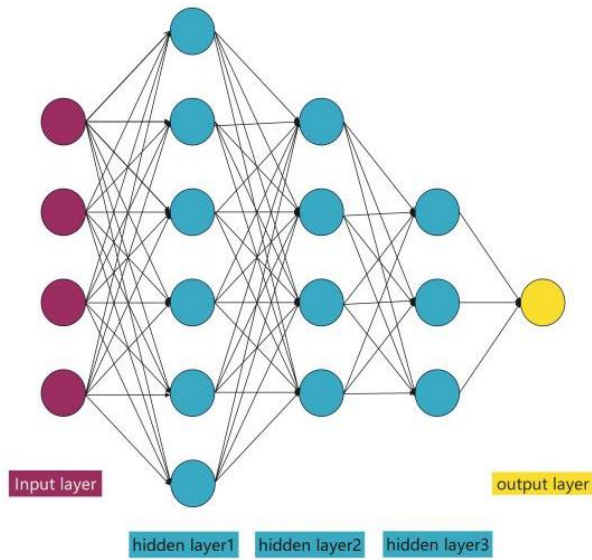

Figure 3. Multilayer neural network structure diagram.

The ANN model used in this study comprises six layers of neurons, with the number of neurons in each layer being 32, 16, 8, 4, 2, and 1, respectively. Except for the final layer, each layer employs the Rectified Linear Unit (ReLU) activation function to enhance the model's capability for non-linear learning. By combining this multi-layer structure with activation functions, the ANN is able to capture complex patterns and relationships within the data, thereby improving the model's predictive performance.

#### 2) Transfer learning

Transfer learning is a machine learning method that enhances performance on a related task by applying knowledge gained from one task (such as model weights) to another shown in Figure 4 Common approaches to transfer learning include feature-based transfer, parameter-based transfer, instance-based transfer, relational transfer, and multi-task learning.

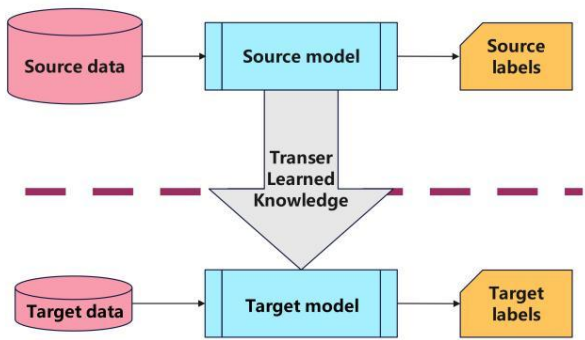

Figure 4. Transfer learning structure diagram.

In this study, the parameter-based fine-tuning method was employed. Initially, the model was trained on a subset with a larger sample size, and the trained model weights were then saved. Subsequently, the weights of the several layers of the model were frozen, and only the remaining layers' weights were fine-tuned on another subset. This approach effectively utilizes the knowledge from prior training to improve training efficiency and model performance on the new data subset.

### D. Implementation details

In this study, the hyperparameter configurations remain consistent whether training on the original subset or the new subset. Specifically, the TensorFlow framework is employed to build the model. The loss function used is Mean Absolute Error (MAE), the optimizer is Adam. For training on the original dataset, the number of epochs is set to 500, whereas for transfer learning on the new subset, it is reduced to 100 epochs. In addition, the batch size is set to 10. During the training process, the evaluation metrics include Root Mean Square Error (RMSE), MAE, and the coefficient of determination ($R^2$). These configurations ensure that the training and evaluation conditions are uniform across different datasets, allowing for a fair comparison of model performance.

### III. RESULTS AND DISCUSSION

#### A. The prediction performance for Mathematics and Portuguese Datasets

This study utilized two distinct datasets from mathematics and Portuguese language courses to train and evaluate the model. The original datasets were divided using the k-means clustering method into cluster0_mat and cluster1_mat for the mathematics dataset, and cluster0_por and cluster1_por for the Portuguese dataset. The model was trained on cluster0 of each dataset and then evaluated on cluster1.

Figure 5 shows the training loss over epochs for the mathematics and Portuguese datasets. In the mathematics dataset, the loss curve gradually decreases and stabilizes, with an initially high loss value that eventually becomes lower. This indicates that the model progressively learns and improves on the mathematics dataset, with a relatively smooth convergence process, ultimately selecting the optimal model weight of 3.6. Conversely, in the Portuguese dataset, the loss curve drops sharply during the initial stages of training and quickly stabilizes, with the final loss value approaching zero. This suggests that the model quickly learns effective features from the Portuguese dataset and further optimizes in later stages, with a faster convergence rate. The model performs exceptionally well on the Portuguese dataset, almost perfectly fitting the training data, and ultimately selects the optimal model weight of 2.4.

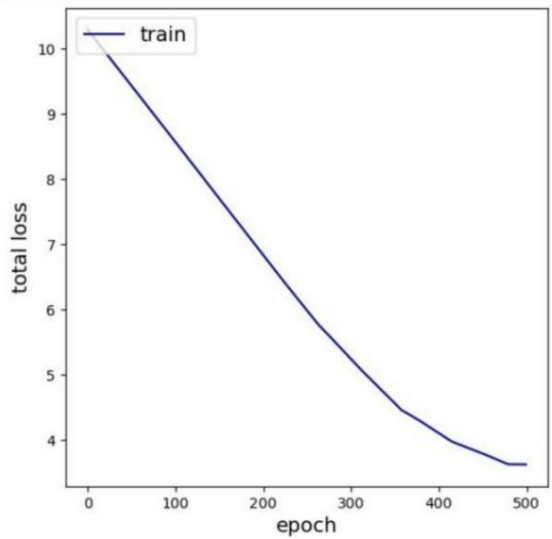

(a) In Mathematics

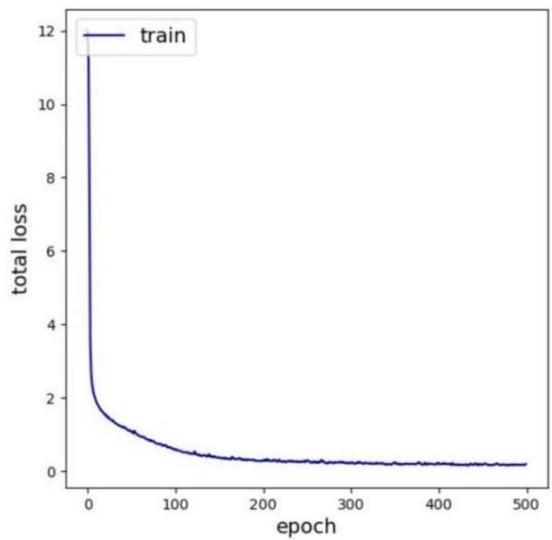

(b) In Portuguese

Figure 5. Training Loss Over Epochs.

Figure 6 presents the relationship between the predicted and real values for the testing dataset of cluster0_mat and cluster0_por. Figure 7 further demonstrates the results of predicting cluster1 using the models directly trained on the mathematics and Portuguese datasets. The RMSE, MAE, and $R^2$ values for these predictions are shown in Table 1.

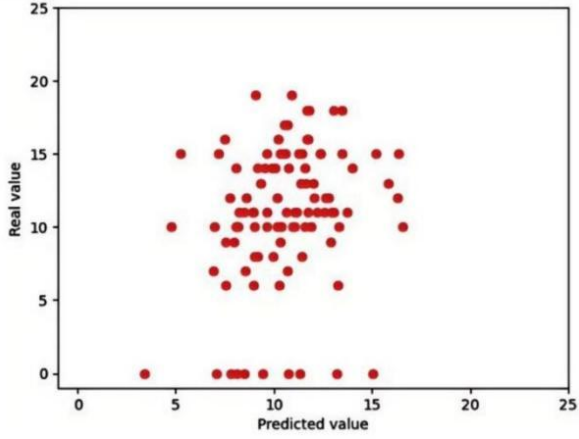

(a) In Mathematics

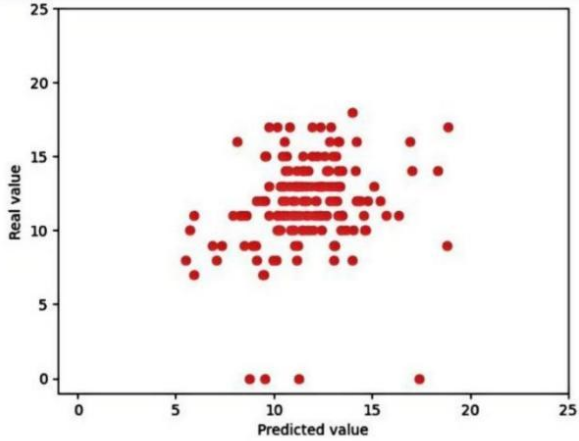

(b) In Portuguese

Figure 6. Scatter Plot of Predicted vs. Real Values in cluster0.

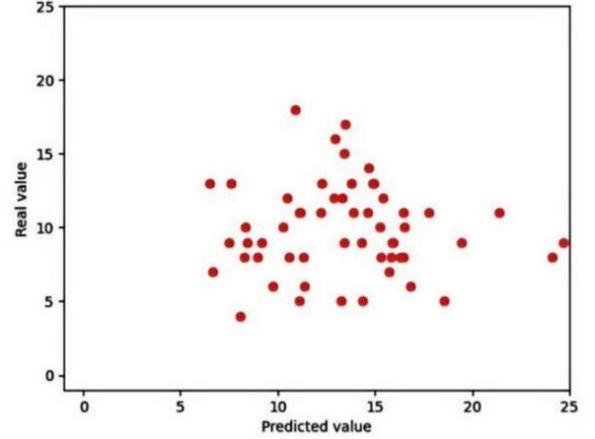

(a) In Mathematics

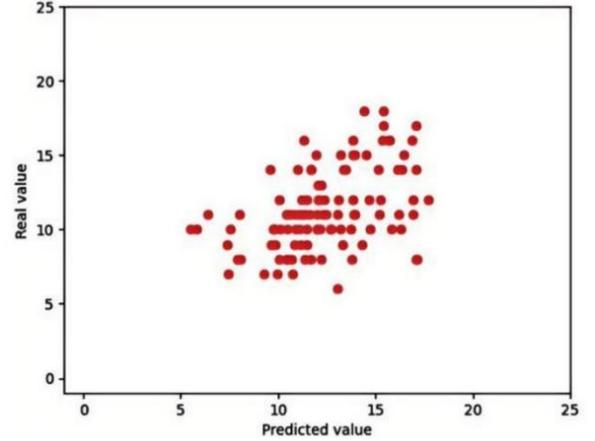

(b) In Portuguese

Figure 7. Scatter Plot of Predicted vs. Real Values in cluster1.

By comparing Figures 6 and 7, it is evident that the model's predictive performance on the testing dataset of cluster0 is significantly better than on cluster1. This observation is further supported by the evaluation metrics in Table 1. For cluster0, the RMSE and MAE values are lower, and the $R^2$ value is relatively higher for both the mathematics and Portuguese datasets, indicating better predictive performance on this subset. However, for cluster1, although the Portuguese dataset maintains relatively low RMSE and MAE values, the mathematics dataset shows significantly higher RMSE and MAE values, and the $R^2$ value is negative, indicating poor predictive performance on this subset.

Comparing the (a) and (b) subplots in Figures 6 and 7, it is clear that the predicted values for the Portuguese dataset have a higher correlation with the real values, and the model performs better on this dataset compared to the mathematics dataset. The predicted values for the mathematics dataset are more dispersed with more outliers, indicating lower prediction accuracy for this dataset.

TABLE I. EVALUATION METRICS FOR UNUSED TRANSFER LEARNING

| Dataset | Evaluation metrics | Math | Por |
|---|---|---|---|
| In cluster0 (Source data) | RMSE | 4.79 | 3.39 |
| | MAE | 3.67 | 2.44 |
| | $R^2$ | -0.03 | -0.24 |
| In cluster1 (Target data) | RMSE | 6.20 | 3.03 |
| | MAE | 4.76 | 2.33 |
| | $R^2$ | -3.11 | -0.25 |

### B. Using Transfer Learning to Improve Prediction Accuracy

To enhance prediction accuracy, this study employed transfer learning techniques. The model, initially trained on cluster0, applied transfer learning by progressively freezing the weights of the first three layers while fine-tuning the remaining layers, undergoing 100 epochs of training.

TABLE II. EVALUATION METRICS FOR USING TRANSFER LEARNING

| Frozen layer | Evaluation metrics | Math | Por |
|---|---|---|---|
| 1 | RMSE | 4.91 | 2.69 |
|   | MAE | 3.85 | 2.15 |
|   | $R^2$ | -1.58 | 0.02 |
| 2 | RMSE | 4.67 | 2.71 |
|   | MAE | 3.71 | 2.16 |
|   | $R^2$ | -1.33 | 0.01 |
| 3 | RMSE | 4.57 | 2.71 |
|   | MAE | 3.65 | 2.17 |
|   | $R^2$ | -1.24 | 0.00 |

Table 2 presents the evaluation metrics RMSE, MAE, and $R^2$ on the test set. For the mathematics dataset, the optimal results were achieved by freezing the third layer, yielding an RMSE of 4.57, MAE of 3.65, and $R^2$ of -1.24. For the Portuguese dataset, the best results were obtained by freezing the first layer, with an RMSE of 2.69, MAE of 2.15, and $R^2$ of 0.02. Figure 8 shows scatter plots of the predicted versus actual values for the optimal frozen layers. In the mathematics dataset, although predictions are somewhat dispersed, there is notable improvement; in the Portuguese dataset, the correlation between predicted and actual values is higher, with the model performing significantly better.

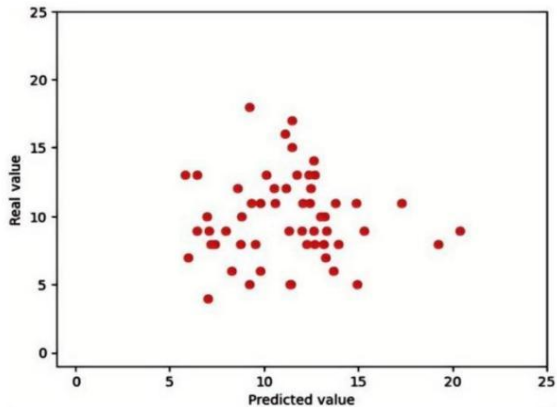

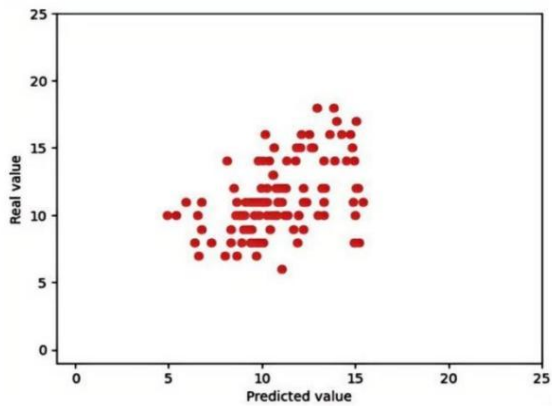

Figure 8. Scatter plot of predicted and true values using transfer learning.

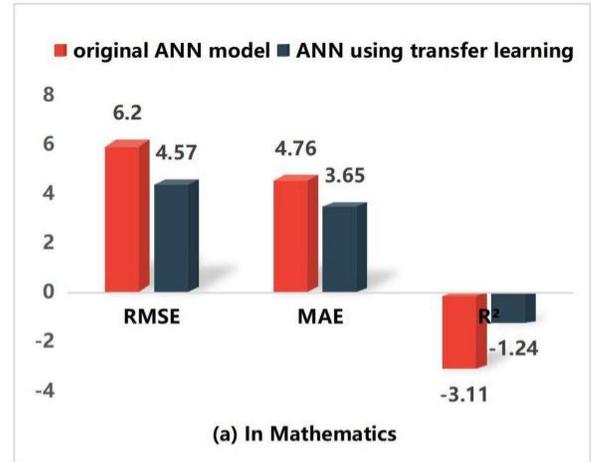

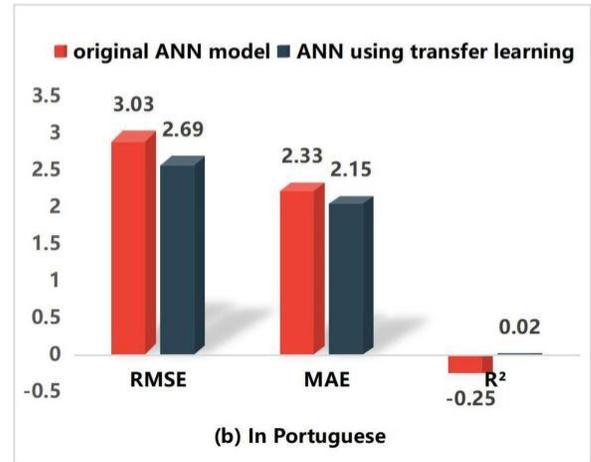

Figure 9. Comparison of evaluation indicators before and after using transfer learning.

Figure 9 compares the results of using transfer learning with those of not using it. The transfer learning approach notably reduces RMSE and MAE values and increases $R^2$ values for both datasets, demonstrating its effectiveness in enhancing model prediction performance.

C. *The influence of freeze layers in performance*

The data in Table 2 further reveal the impact of freezing different numbers of layers on model performance. For the mathematics dataset, freezing more layers (e.g., the third layer) results in better predictive performance, possibly due to the dataset's complexity and higher noise, which require more prior knowledge to improve generalization. Conversely, for the Portuguese dataset, freezing fewer layers (e.g., the first layer) significantly enhances performance, indicating that the dataset is simpler, and the model can quickly learn and effectively utilize the previously trained knowledge. The amount of target domain data also influences the effect of frozen layers. The Portuguese dataset has a larger data volume (122 entries), whereas the mathematics dataset has fewer entries (64 entries).

When the data volume is small, freezing more layers helps prevent overfitting and enhances generalization. With larger data volumes, freezing fewer layers allows the model to adequately train on new data, improving performance. In summary, transfer learning, by freezing some layers and fine-tuning the weights, significantly improves the model's predictive accuracy across different datasets.

IV. CONCLUSION

This study proposes a method combining transfer learning to improve the accuracy of student performance predictions under different data distributions. Using datasets from mathematics and Portuguese courses, the model significantly enhanced its generalization ability and prediction accuracy by freezing some layer weights and fine-tuning others, particularly excelling in RMSE and MAE metrics. Experimental results indicate that transfer learning effectively improves the model's performance across different datasets. Research on the impact of freezing layers suggests that the size of the target domain data and the complexity of the data may influence the choice of which layers to freeze. Freezing more layers helps prevent overfitting when the data volume is small or the data complexity is high, while freezing fewer layers allows for better utilization of new data when the data volume is large and the data complexity is low. Although this study has achieved promising results, the current method relies on labeled target datasets. Future research could explore domain adaptation techniques for unlabeled datasets and integrate other improvement methods to further enhance the practical applicability of the model.

REFERENCES

[1] B. Sravani and M. M. Bala, "Prediction of student performance using linear regression," in Proc. 2020 International Conference for Emerging Technology (INCET), Jun. 5, 2020, pp. 1-5.
[2] V. A. Ramesh, P. Parkavi, and K. Ramar, "Predicting student performance: a statistical and data mining approach," International Journal of Computer Applications, vol. 63, no. 8, Jan. 1, 2013.
[3] Y. Qiu, Y. Hui, P. Zhao, C. H. Cai, B. Dai, J. Dou, S. Bhattacharya, and J. Yu, "A novel image expression-driven modeling strategy for coke quality prediction in the smart cokemaking process," Energy, vol. 294, p. 130866, May 1, 2024.
[4] P. Fang, W. Zecong, and X. Zhang, "Vehicle automatic driving system based on embedded and machine learning," in Proc. 2020 International Conference on Computer Vision, Image and Deep Learning (CVIDL), Jul. 10, 2020, pp. 281-284.
[5] Y. Qiu, J. Wang, Z. Jin, H. Chen, M. Zhang, and L. Guo, "Pose-guided matching based on deep learning for assessing quality of action on rehabilitation training," Biomedical Signal Processing and Control, vol. 72, p. 103323, Feb. 1, 2022.
[6] S. Li, G. Wei, and Z. Zhao, "A Method for Predicting Comprehensive Grades of College Students Based on Multi Model Fusion," Control Engineering, pp. 1-8, May 19, 2024. [Online]. Available: https://doi.org/10.14107/j.cnki.kzgc.20240049
[7] A. Kukkar, R. Mohana, A. Sharma, et al., "A novel methodology using RNN+ LSTM+ ML for predicting student's academic performance," Education and Information Technologies, pp. 1-37, 2024.
[8] A. Durak and V. Bulut, "Classification and prediction-based machine learning algorithms to predict students' low and high programming performance," Computer Applications in Engineering Education, vol. 32, no. 1, p. e22679, 2024.
[9] L. A. Shoaib, S. H. Safii, N. Idris, et al., "Utilizing decision tree machine model to map dental students' preferred learning styles with suitable instructional strategies," BMC Medical Education, vol. 24, no. 1, p. 58, 2024.
[10] Y. Manzali, Y. Akhiat, K. Abdoulaye Barry, et al., "Prediction of Student Performance Using Random Forest Combined With Naïve Bayes," The Computer Journal, p. bxae036, 2024.
[11] M. Laurer, W. van Atteveldt, A. Casas, and K. Welbers, "Less Annotating, More Classifying: Addressing the Data Scarcity Issue of Supervised Machine Learning with Deep Transfer Learning and BERT-NLI," Political Analysis, vol. 32, no. 1, pp. 84-100, 2024. doi:10.1017/pan.2023.20
[12] Y. Ma, S. Chen, S. Ermon, et al., "Transfer learning in environmental remote sensing," Remote Sensing of Environment, vol. 301, p. 113924, 2024.
[13] B. M. Ampel, S. Samtani, H. Zhu, et al., "CREATING PROACTIVE CYBER THREAT INTELLIGENCE WITH HACKER EXPLOIT LABELS: A DEEP TRANSFER LEARNING APPROACH," MIS Quarterly, vol. 48, no. 1, 2024.
[14] Kaggle, "Student Performance Prediction," Available: https://www.kaggle.com/datasets/henryshan/student-performance-prediction.
[15] M. Ahmed, R. Seraj, and S. M. S. Islam, "The k-means algorithm: A comprehensive survey and performance evaluation," Electronics, vol. 9, no. 8, p. 1295, 2020.
[16] C. Shi, B. Wei, S. Wei, et al., "A quantitative discriminant method of elbow point for the optimal number of clusters in clustering algorithm," Eurasip Journal on Wireless Communications and Networking, vol. 2021, no. 1, p. 1-16, 2021.
[17] A. Maćkiewicz and W. Ratajczak, "Principal components analysis (PCA)," Computers & Geosciences, vol. 19, no. 3, pp. 303-342, 1993.